\documentclass[letter]{aa} 
\usepackage{graphicx}
\usepackage{txfonts}
\begin{document}

    \title{Silicate features in the circumstellar envelopes of the Class~I binary driving source of HH~250\thanks{Based on observations collected at the European Organisation for Astronomical Research in the Southern Hemisphere under ESO programme 0101.C-0126(A).}
   }

   \author{F. Comer\'on\inst{1}
          \and
          B. Mer\'\i n \inst{2}
          \and
          B. Reipurth\inst{3}
          \and
          H.-W. Yen \inst{1}
          }

   \institute{European Southern Observatory, Karl-Schwarzschild-Strasse 2, D-85748 Garching bei M\"unchen, Germany\\
              \email{fcomeron@eso.org}
         \and
   {European Space Astronomy Centre (ESAC), European Space Agency (ESA), E-28691 Villanueva de la Cañada, Spain}
   \and
   {Institute for Astronomy, University of Hawaii at Manoa, 640 N. Aohoku Place, Hilo, HI 96720, USA}
 }

   \date{Received; accepted}


  \abstract
   {The silicate feature near 10~$\mu$m is one of the main tools available to study the mineralogy of circumstellar disks and envelopes, providing information on the thermal processing, growth, location, and circulation of dust grains.}
   {We investigate the silicate feature of the two Class~I components of HH~250-IRS, a resolved binary system with a separation of $0''53$ driving a Herbig-Haro flow. Each component has its own circumstellar envelope, and the system is surrounded by a circumbinary disk.}
   {We have carried out low resolution spectroscopy in the 8-13~$\mu$m range using VISIR, the thermal infrared imager and spectrograph at ESO's Very Large Telescope.}
   {The silicate features of both sources are clearly different. The NW component has a broad, smooth absorption profile lacking structure. We attribute most of it to foreground interstellar dust absorption, but estimate that additional absorption by amorphous silicates takes place in the circumstellar envelope of the young stellar object. The SE component shows the silicate feature in emission, with structure longwards of $9.5$~$\mu$m indicating the presence of crystalline dust in the dominant form of forsterite. The apparent lack of an absorption feature caused by foreground dust is probably due to the filling of the band with emission by amorphous silicates in the envelope of the object.}
   {Despite their virtually certain coevality, the differences in the components of the HH~250-IRS binary are most likely due to markedly different circumstellar environments. The NW component displays an unevolved envelope, whereas dust growth and crystallization has taken place in the SE component. The weak or absent signatures of enstatite in the latter are fairly unusual among envelopes with crystalline dust, and we tentatively relate it to a possible wide gap or an inner truncation of the disk already hinted in previous observations by a drop in the $L'$-band flux, which might indicate that the SE component could actually be a very close binary. We speculate that the clear differences between the silicate feature spectra of both components of HH~250-IRS may be due either to disk evolution sped up by multiplicity, or by accretion variability leading to episodes of crystal formation. Different inclinations with respect to the line of sight may play a role as well, although it is very unlikely that they are the sole responsible for the differences between both objects.}

   \maketitle
%

\section{Introduction}

Mid-infrared spectroscopy of young stellar objects provides important clues about their circumstellar envelopes and disks through the appearance of emission and absorption features that make it possible to study their mineralogy. The intensity and structure of those features reveal a wealth of information about aspects such as the composition of the dust, its growth, the location where its composing mineral species are formed, and their thermal history. The study of those features provides an extremely valuable link between the processes of dust formation in the winds of red giant stars, its presence and conditions in the surroundings of young stellar objects, the formation of planetary systems around other stars, and also about our own Solar System where comets are carriers of pristine material 

HH~250-IRS (=IRAS~19190+1048), the source that powers the Herbig-Haro object HH~250, is a binary system composed of two Class~I young stellar objects separated by $0''53$. The system lies in the outskirts of the Aquila Rift \citep{Prato08} and is embedded in the LDN~643 dark cloud. An investigation of the young stellar population in the Serpens-Aquila region lying in the same general direction of the sky using the Gaia DR2 catalog \citep{GaiaDR2} has revealed an aggregate of at least 15 young stars associated with LDN~643, including the previously identified binary T~Tauri stars AS~353 and V536~Aql \citep{Herczeg19}. Astrometry of the members of the aggregate yields a distance of $407 \pm 16$~pc, which is very similar to that of other aggregates in the Aquila Rift and the Serpens star forming clouds.
Adopting this distance for HH~250-IRS translates into a projected separation of 216~AU\footnote{The Gaia DR2 distance is almost twice the value adopted in \citet{Comeron18}, before the Gaia DR2 data became available. Some of the quantities discussed in that article need to be rescaled using the new distance, but the results obtained in the present work are independent of the distance.} between its components. Observations in the visible, infrared and submillimeter reported in \citet{Comeron18} (hereafter CRYC18) show that each of the binary components is surrounded by large amounts of circumstellar material and that the binary is itself surrounded by a marginally resolved circumbinary disk. HH~250 and HH~250-IRS are thus one of the rare examples of young stellar objects simultaneously displaying features of early stellar evolution such as an active outflow, a circumbinary disk, and two circumstellar disks that can be studied separately thanks to subarcsecond resolution instrumentation currently available at near-infrared and submillimeter wavelengths.

Despite similarities in the shape of the largely featureless near-infrared spectra of both sources presented by CRYC18, some clear differences appear at longer wavelengths. One of the components, HH~250-IRS SE, has overall bluer colors than the NW component, a marked drop in flux at 3.8~$\mu$m, and hints of an emission component in the $N$ band possibly due to a silicate feature in emission. The NW component has a redder and smoother spectral energy distribution, in this case with hints of a silicate feature in absorption. Despite their similarities, small separation and coevality, the spectral energy distribution traced by the broad-band photometry presented in CRYC18 suggests substantial differences between the circumstellar envelopes of both objects.

In this Letter we report the result of spectroscopic observations in the 10~$\mu$m atmospheric window exploring the silicate features of both sources. Besides confirming preliminary findings by CRYC18, we further discuss evidence for dust processing in the SE component, where evidence of crystallinity is found.

\section{Observations\label{observations}}

The observations reported here were obtained with VISIR \citep{Lagage04}, the thermal infrared imager and spectrograph at ESO's Very Large Telescope (VLT), in spectroscopic mode. A low-resolution grating was used covering the 8~$\mu$m to 13~$\mu$m interval at a resolution $R = 270$ with the $0''4$ slit used. The slit was oriented so as to include both components of the binary in every exposure. A combination of chopping and nodding along the slit direction with an amplitude of $8''$ was used to cancel out the sky emission as well as the signature due to differences in the optical path when chopping.

The observations were obtained on the night of 22 to 23 June 2018. A total of 18 nodding cycles, each with 56 chopping cycles of the telescope secondary mirror at a frequency of 0.8~Hz, were performed. The total exposure time was 2140~s. However, approximately 20\% of the set of exposures taken at given nod positions were rejected due to transient image quality degradations that caused overlap between the traces of the spectra of both components of the binary system. The K5III star HD 189695 was observed at a very similar airmass right after the exposures on our science targets to allow the removal of telluric features.

\begin{figure}[ht]
\begin{center}
\hspace{-0.5cm}
\includegraphics [width=8cm, angle={0}]{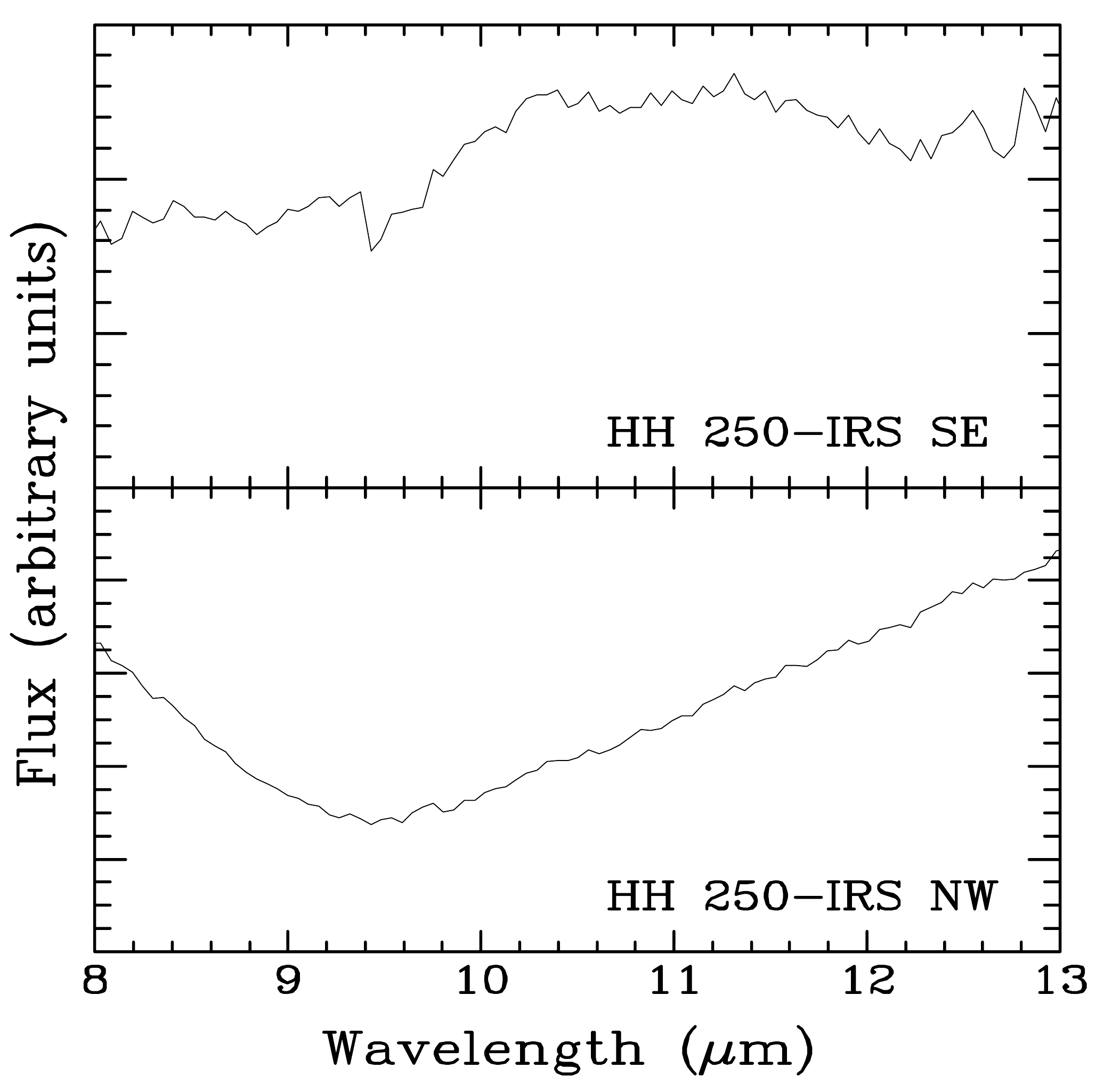}
\caption []{Spectra of both components of the HH~250-IRS binary in the 8-13~$\mu$m region. The spectra are smoothed to the resolution $\lambda / \Delta \lambda \simeq 170$ of the flux-calibrated spectrum of the telluric correction star \citep{Cohen99}.}\label{spec}
\end{center}
\end{figure}

Standard IRAF tasks were used for data reduction and calibration. Spectra at each of the two positions along the slit were extracted and stacked together for each target source and for the telluric correction star. The trace of the telluric correction spectra was used to extract a spectrum of the sky at each of the two slit positions, which was used for wavelength calibration. The stacked, wavelength calibrated spectra of the target object at each of the two slit positions were then combined, and the same was done for the telluric correction star. Telluric features removal was carried out by dividing the spectra of the target stars by that of the telluric correction star, and the result was multiplied by a template spectrum of the K3III star $\alpha$~Hya \citep{Cohen99} in order to get a relative flux calibration. The resulting spectra are shown in Figure~\ref{spec}, where differences between both sources are obvious.

\section{Results\label{results}}

The advent of mid-infrared spectrographs onboard space observatories such as ISO, Spitzer and Herschel, and of sensitive mid-infrared spectrographs at large ground-based telescopes, has produced large collections of spectra with silicate features around young stellar objects of various masses, ranging from Herbig Ae/Be stars \citep{vanBoekel05,Sicilia11} down to lower mass young stellar objects \citep{Kessler05}, and all the way to the substellar limit \citep{Merin07}, illustrating a rich diversity that can be interpreted in terms of mixtures of different species, grain sizes, locations, and processes of crystal formation. Such diversity is found even in our own planetary system, as illustrated by the silicate features observed in cometary comae \citep{Wooden07}.

\subsection{HH~250-IRS~NW and the contribution of interstellar extinction\label{NW}}

The $8-13$~$\mu$m spectrum of HH~250-IRS~NW (Figure~\ref{spec}, bottom panel) shows a smooth, structureless feature in absorption with a minimum at 9.4~$\mu$m. The feature has the usual shape produced by small ($\sim 0.1$~$\mu$m) grains of amorphous olivine and pyroxene silicate. The feature may be due either to absorption by unprocessed dust in the circumstellar envelope, or to interstellar grains in the intervening dust column. Images at visible wavelengths of the region where both sources are embedded (see Figure~1 of CRYC18) clearly show that there is a considerable amount of extinction toward them, which should be accounted for at the time of discussing the intrinsic properties of the circumstellar environments of the HH~250-IRS binary components.

In Figure~4 of CRYC18 we presented the $L'$-band spectrum toward both sources, which shows a prominent, broad absorption feature centered around 3.1~$\mu$m due to interstellar water ice grains. By a rough extrapolation of the continuum longward of the feature we estimate an optical depth $\tau_{\rm H_2O} \simeq 0.6$ at the core of the feature, which corresponds to a visible extinction $A_V \simeq 11.5$~mag \citep{Whittet01}. On the other hand, we estimate $\tau_{\rm Si} \simeq 0.85$ at the core of the silicate feature in the spectrum of HH~250-IRS~NW by interpolating between the fluxes at 8~$\mu$m and 13~$\mu$m. This is most likely an underestimate, since the edges of the feature extend somewhat beyond both limits and the interpolated continuum should actually be higher than the level that we have assumed. The contribution of the foreground interstellar dust to the feature is estimated to be $\tau_{\rm Si, ISM} \simeq A_V / 16.6 \simeq 0.7$ using the $A_V$-to-$\tau_{\rm Si, ISM}$ proportionality of \cite{Rieke85} and the value of $A_V$ derived from the depth of the 3.1~$\mu$m ice feature. Interstellar dust absorption thus seems to be insufficient to account for the full depth of the silicate feature in the spectrum of HH~250-IRS~NW, leading us to conclude that the intrinsic spectrum of its envelope contains the signature of amorphous silicate absorption.

\subsection{HH~250-IRS~SE\label{SE}}

No obvious feature of the same shape as the one dominating the $8-13$~$\mu$m spectrum of HH~250-IRS~NW appears in the spectrum of the SE component (Figure~\ref{spec}, upper panel). However, being just $0''53$ from the NW component and therefore equally embedded in the molecular cloud, the amount of foreground extinction is expected to be essentially identical. The most likely cause of its absence in the spectrum of HH~250-IRS~SE is the existence of emission by the same species in its circumstellar envelope, with an intensity that approximately fills in the absorption caused by foreground dust. We will assume henceforth this to be the case.

\begin{figure}[ht]
\begin{center}
\hspace{-0.5cm}
\includegraphics [width=9cm, angle={0}]{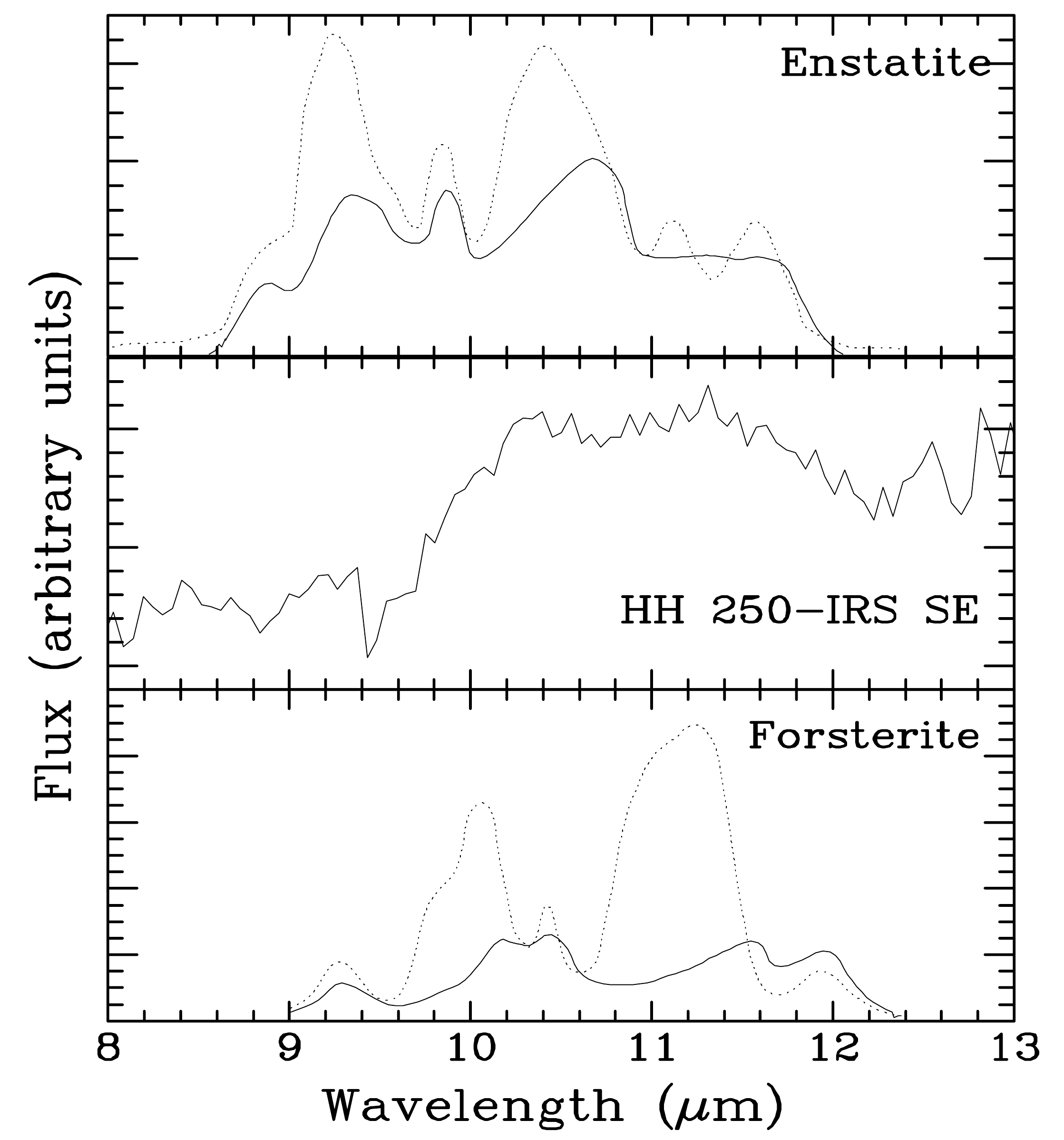}
\caption []{Silicate feature of HH~250-IRS~SE (middle panel) compared to the spectra of enstatite (top panel) and forsterite (bottom panel) The spectra of crystalline silicates are adapted from \citet{vanBoekel05}. To facilitate the comparison a constant level has been subtracted from the spectrum of HH~250-IRS~SE. The solid and dotted curves in the spectra of both silicates correspond to grains of $1$ and $0.1$~$\mu$m, in size, respectively.}\label{spec_comp}
\end{center}
\end{figure}

The most obvious characteristic of the spectrum of HH~250-IRS~SE is the flux increase that dominates the longer wavelength part of the interval shown in Figure~\ref{spec}. Such a structured feature indicates the presence of emission due to crystalline dust species. The flux increase starts around 9.5~$\mu$m and reaches a plateau extending up to $11.2$~$\mu$m. Figure~\ref{spec_comp} compares the structured emission feature with the mass absorption coefficients of the two most common crystalline silicates in circumstellar disks, enstatite (MgSiO$_3$) and forsterite (Mg$_2$SiO$_4$), for two typical grain sizes of each species \citep{vanBoekel05}.

The main characteristic of the feature, namely the rise in flux between $\sim 9.5$ and $\sim
10.2$~$\mu$m, can be best reproduced with crystalline grains of forsterite as a main component. No emission features attributable to enstatite are obvious shortward of 9.5~$\mu$m, whose most prominent feature appears near 9.4~$\mu$m \citep{Koike00,Molster02,Olofsson09}, indicating that forsterite is the dominant form of crystalline silicates in the circumstellar enviroment of HH~250-IRS~SE. As can be clearly seen in the comparison spectra of Figure~\ref{spec_comp}, the bands characteristic of the spectrum of forsterite become decreasingly prominent with increasing grain size, and dilute into a continuum above grain sizes $\sim 1.6$~$\mu$m \citep{Lindsay13}. The fact that the signature of forsterite is still visible in our spectrum, although not in the form of well-defined peaks, indicates the dominance of rather large grains and that significant grain growth through coagulation has taken place, up to sizes near $\sim 1$~$\mu$m. On the other hand, \citet{Lindsay13} have investigated the dependency of the features on the shapes of the forsterite crystals, which in turn is related to their temperature of formation \citep{Kobatake08}. In the shape classification of \citet{Lindsay13}, both {\it a-platelets} and {\it c-columns} have the dominant 10~$\mu$m feature shifted to a peak at 10.2~$\mu$m, like what we see in the spectrum of HH~250-IRS~SE. {\it c-columns} have the 11~$\mu$m feature shifted to wavelengths shorter than 11~$\mu$m, which we do not see in our spectra. We therefore interpret the crystalline features that we observe in HH~250-IRS~SE as being predominantly due to micron-size, perhaps {\it a-platelet}-shaped grains of forsterite, although the latter conclusion is rather tentative and should be confirmed with higher signal-to-noise spectroscopy.

\section{Discussion}

All the characteristics of HH~250-IRS~NW noted in CRYC18 and here tend to confirm its classification as a relatively unevolved Class~I object, surrounded by a thick envelope where dust is predominantly in the amorphous state. Since evidence for significant dust grain growth exists already during the relatively short-lived Class~I phase \citep{Sheehan17}, it is likely that it has taken place in HH~250-IRS~NW already to some degree, but without producing yet detectable amounts of crystallinity, which is a slower process than coagulation \citep{Suttner01}. An alternatively possibility might be that crystalline evolution has taken place in the inner disk of HH~250-IRS~NW, but that the disk is seen close to edge-on from our vantage point, resulting in an optically thick outer layer of amorphous dust blocking our view of the central regions. This seems nevertheless unlikely, as computed model spectral energy distributions predict a much deeper flux drop, by several orders of magnitude, in the 10~$\mu$m region \citep{Robitaille06} for objects in early evolutionary stages, even when the geometry departs from an edge-on orientation by a few tens of degrees.

Given the close proximity between the two components of HH~250-IRS, it is possible that the overall evolution of their circumstellar envelopes has been accelerated by the presence of the companion. Using submillimeter observations of a sample of Class~0 and Class~I binaries in Ophiuchus with separations in the range 450-1100~AU, \citet{Patience08} have shown that the frequency of circumstellar envelopes around the secondary component of the system drops abruptly from Class~0 to Class~I. Their submillimeter observations are mainly sensitive to the evolution of the outer parts of envelopes, but demonstrate that the presence of a companion causes substantial evolution of the envelope even on timescales of $\sim 10^5$~yr or shorter. Such influence can be expected to be even more dramatic in the case of HH~250-IRS, in which the separation between components is between 2 and 5 times smaller than those probed by those authors. In agreement with the results of \citet{Patience08}, CRYC18 find only one compact source of dust continuum emission in the HH~250-IRS binary. The position of that source seems to be closer to HH~250-IRS~SE, although this should be confirmed by more astrometrically accurate observations.

The presence of forsterite as the dominant crystalline species is a remarkable characteristic of the spectrum of HH~250-IRS~SE, as the features of both species are most often observed simultaneously in the spectra of circumstellar environments. Forsterite formed by thermal annealing is the main form of silicate crystals in the innermost regions of circumstellar disks \citep{vanBoekel05}, reproducing model expectations \citep{Gail04}. Enstatite dominates the inner warm regions of the disk, but further outwards forsterite is found to be again the dominant form of crystalline silicates \citep{Bouwman08}, in contrast with model predictions, perhaps indicating more complicated heating processes not considered in stationary disk models \citep{Henning10}, or as a consequence of the slow forsterite to enstatite conversion at low temperatures \citep{Gail04}.

We tentatively link the dearth of enstatite with the lack of dust in the inner disk that CRYC18 proposed as an explanation for the drop in $L'$-band flux of the spectral energy distribution of HH~250-IRS~SE. The dust-devoid region may encompass the range of distances to the central star where enstatite would otherwise be the predominant crystalline silicate. The clearing of dust might be triggered by tidal interaction with HH~250-IRS~NW. However, simulations of the interaction between circumstellar disks and binary companions at distances of several tens or hundreds of AU, which result in truncation of the outer disk and tidal tails, do not seem to be able to reproduce such a feature \citep{Artymowicz94}. This has been observationally supported by \citet{Pascucci08}, who found no statistical difference in the degree of crystallinity and dust settling between a sample of single T~Tauri stars, and another composed of binaries with separations between 10~AU and 450~AU, which suggests that the presence of companions in that range does not significantly affect the inner disk. An intriguing alternative possibility is that HH~250-IRS~SE may itself be a tight binary system, in which the tidal interaction with the central binary is responsible for the inner truncation of the disk common to both components. \citet{Andrews10} have studied two such cases in the TW~Hya association, and \citet{Ruiz16} have shown that over one third of a sample of young stellar objects in nearby star forming regions with spectral energy distributions typical of transitional disks, lacking near-infrared excesses, are actually close binaries. If this were the case of HH~250-IRS~SE as well, then HH~250-IRS would be a hierarchical triple system with two nested circumbinary disks, plus one circumstellar disk around the NW component. This is the expected outcome of the dynamical decay of a non-hierarchical triple system initially embedded in the dense interior of a cloud core \citep[e.g.][]{Reipurth10,Reipurth15}. Such explanation is nevertheless speculative at this point in the absence of any other evidence of close binarity of HH~250-IRS~SE.

The match between the structure of the crystalline silicate feature and the spectral characteristics of {\it a-platelets} also provides information on the temperature of formation. \citet{Kobatake08} show that {\it a-platelets} are preferentially formed at temperatures between 970~K and 1270~K, with more symmetric structures \citep[equants in the terminology of][]{Lindsay13} formed at the highest temperatures and  columnar structures formed below. As noted by \citet{Lindsay13}, the spectral signatures of platelet shapes are not detected in comets, implying that cometary dust in our Solar System condensed at higher temperature.

In view of the virtually certain coevality of the two components of the binary, the results obtained in CRYC18 and enlarged here by means of spectroscopy of the silicate feature caution against an interpretation of crystallinity as a straightforward indicator of the evolutionary stage of a circumstellar disk. SVS~20, another Class~I binary discovered by \citet{Eiroa87}, has some features in common with HH~250-IRS, displaying similar spectral energy distributions with different mineralogy in the 10~$\mu$m band \citep{Ciardi05}. However, the spectral energy distributions of both components of SVS~20 are much more similar to each other, and both components exhibit the crystalline emission features that are lacking in HH~250-IRS~NW.

The strikingly different characteristics of the dust around the components of HH~250-IRS might be due to variations in the evolutionary timescales of circumstellar envelopes caused by the presence of companions, which in the case of HH~250-IRS-SE may either be the NW component or a hypothetical close companion responsible for producing a large gap in the inner disk. However, other causes are possible. \cite{Abraham09} have shown evidence for fast formation of crystalline features during the 2008 outburst of EX~Lup, with strong indirect evidence that such features could disappear from the spectrum in a matter of decades, most likely due to radial transport of the freshly-formed crystals \citep{Juhasz12}. The existence of the Herbig-Haro object HH~250, with a kinematical age estimate of $\sim 3,500$~years, shows that at least one of the members of the HH~250-IRS system is at the jet-launching phase that usually coincides with the evolutionary stage at which FU~Ori or EX~Lup outbursts take place. Accretion variability, rather than steady disk evolution, could therefore be responsible for the derived crystallinity of HH~250-IRS-SE.

Since the depth of the 10~$\mu$m feature is a sensitive function of the disk orientation along the line of sight, it cannot
be ruled out that non-coplanarity of the disks of both components also plays a role in the different appearance of their spectra. However, the moderate depth of the 10~$\mu$m feature in both objects indicates that neither of them is seen close to edge-on, as discussed above. We thus deem very unlikely that the possibly different orientations of their disks is a dominant factor causing the sharp differences between their spectra.

Our results on the silicate features study adds a new aspect to the wealth of structures and phenomena related to the earliest stages of stellar evolution that are found in the HH~250-IRS system, and presents it as a case study where the effects of multiplicity and accretion variability on the evolution of the mineralogy of the circumstellar environment may be investigated. Further observations, and particularly adaptive optics-assisted photometric and spectroscopic monitoring able to separately follow the evolution of each component, will be helpful in this regard.

\begin{acknowledgements}

We are grateful to Christian Hummel, from ESO's User Support Department, for the careful verification of the Observation Blocks composing our programme, and to the ESO staff astronomers in charge of Service Mode observations at the VLT when our programme was executed. Comments by the anonymous referee greatly helped with the discussion of our results and are gratefully acknowledged.

\end{acknowledgements}

\bibliographystyle{aa}
\bibliography{hh250_spec_cit}

\end{document}